\documentclass[reprint,
superscriptaddress,
amsmath,amssymb,
aps,
prl,
]{revtex4-2}
\usepackage{graphicx}
\usepackage{dcolumn}
\usepackage{bm}
\usepackage{dsfont}
\usepackage{physics}

\begin{document}


\title{Theory of Vibrational Polariton Chemistry in the Collective Coupling Regime}
\author{Arkajit Mandal}%
\email{amandal4@ur.rochester.edu}
\affiliation{Department of Chemistry, University of Rochester, Rochester, New York, 14627}
\author{Xinyang Li}%
\affiliation{Department of Chemistry, University of Rochester, Rochester, New York, 14627}
\author{Pengfei Huo}
\email{pengfei.huo@rochester.edu}
\affiliation{Department of Chemistry, University of Rochester, Rochester, New York, 14627}
\affiliation{Institute of Optics, Hajim School of Engineering and Applied Sciences, University of Rochester, Rochester, New York, 14627}


\date{\today}
\begin{abstract}
We theoretically demonstrate that chemical reaction rate constant can be significantly suppressed by coupling molecular vibrations with an optical cavity, exhibiting both the collective coupling effect and the cavity-frequency modification of the rate constant. When a reaction coordinate is strongly coupled to the solvent molecules, the reaction rate constant is reduced due to the dynamical caging effect. We demonstrate that collectively coupling the solvent to the cavity can further enhance this dynamical caging effect, leading to additional suppression of the chemical kinetics. This effect is further amplified when cavity loss is considered.
\end{abstract}

\maketitle
\begin{figure}
 \centering
  \begin{minipage}[h]{1.0\linewidth}
     \centering
     \includegraphics[width=\linewidth]{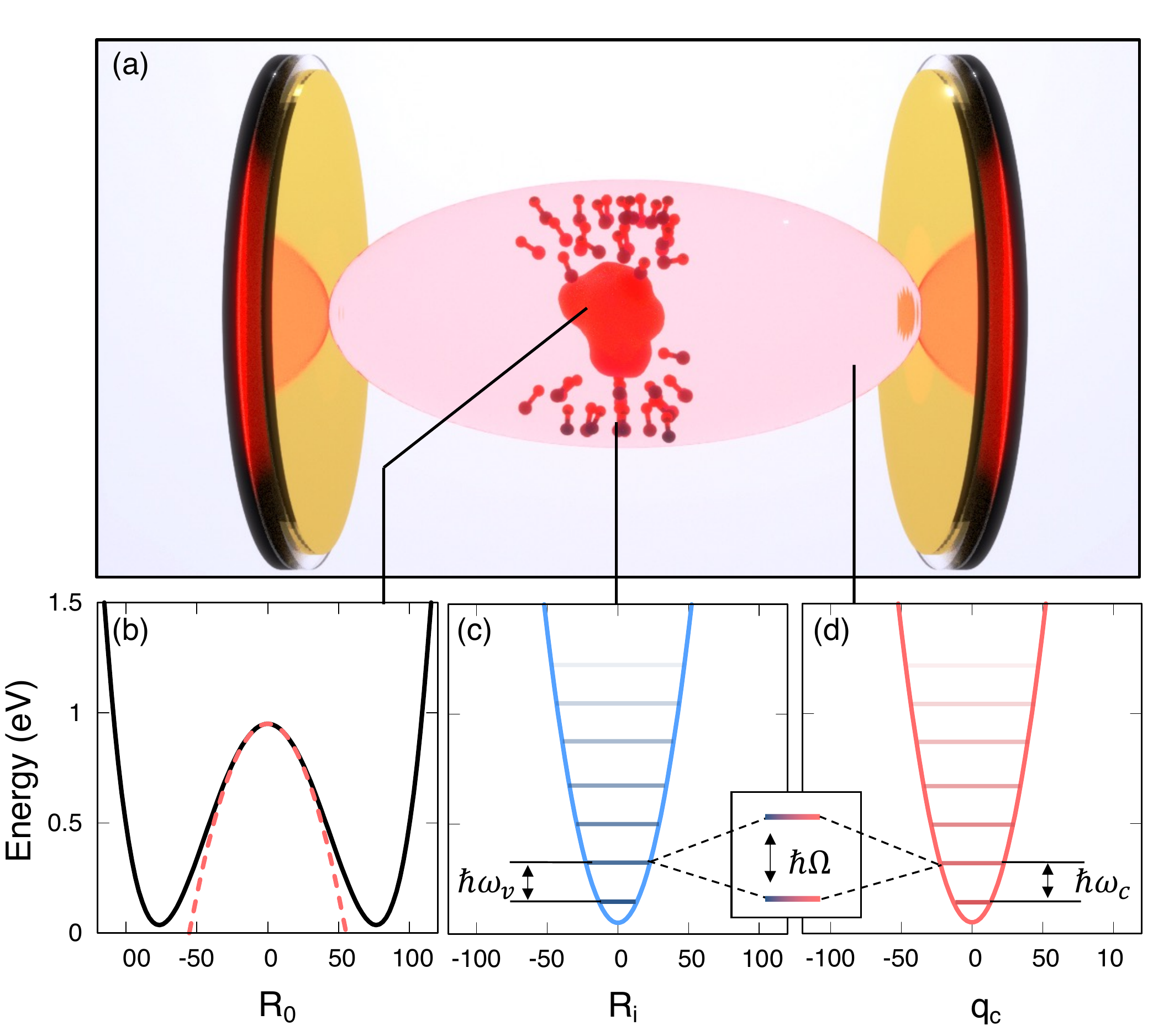}
       \end{minipage}%
   \caption{\small (a) Schematic representation of the reactive molecule coupled to a set of solvent (phonon) modes which are coupled to quantized radiation in a cavity. (b) Double-well potential (black solid line) representing a chemical reaction as a function the reaction coordinate $R_{0}$ where the barrier can approximated as a inverted Harmonic potential (red dash). Potential energy of a (c) Harmonic phonon mode (solvent)  and (c) a bare cavity mode which when hybridized leads to a Rabi-splitting $\hbar \Omega_{R}$.  }
\label{Model}
\end{figure}

Hybridizing molecular vibrations and the photonic excitations inside an optical cavity~\cite{Ebbesen16PRL, Shalabney2015, Ebbesen2019} forms vibrational polaritons (Fig.~\ref{Model}a). Several recent experiments~\cite{Thomas2016, Ebbesen2019, Anoop2020, Vergauwe2019, Hirai2020-Prins, Lather2019} have demonstrated that it is possible to modify ground-state chemical reactions by coupling the cavity radiation mode with the vibrational degrees of freedom (DOF) of molecules. This new strategy of vibrational strong coupling (VSC), if feasible, will offer a paradigm-shift in chemical transformations~\cite{Ebbesen2019, Anoop2020}. Despite recent theoretical works~\cite{li2021collective, DU2021can, Li2021, FeistPCCP, joel2020, TaoJCP2020, schfer2021shining}, a clear theoretical explanation of such remarkable VSC effects in ground-state reactivity remains elusive, including explanations of both (i) the collective effect ($N$-dependent effect where $N$ is the total number of molecules inside the cavity) on chemical reaction rates and (ii) the resonant effect where the suppression of the rate is achieved with a particular cavity photon frequency. In this work, we theoretically demonstrate that the chemical reaction rate constant can be modified when a set of solvent DOF is collectively coupled to both a reaction coordinate of a solute molecule and the cavity radiation mode. Our results demonstrate both the collective coupling effect and the cavity frequency-dependent modifications of the reaction rate constant, which purely originate from a change of the transmission coefficient (recrossing factor) of the rate constant due to the dynamical caging effects from the cavity~\cite{Li2021}. 

We begin by writing the light-matter interaction Hamiltonian in the minimal coupling form as follows
\begin{equation}\label{eqn:Hc}
\hat{H}_\mathrm{C}=\sum_{j}\frac{1}{2m_j}(\hat{\bf p}_{j}-{z}_j\hat{\bf A})^2+\hat{V}(\hat{\bf x})+\hat{H}_\mathrm{ph},
\end{equation}
where the sum is performed over all charged particles, including both electrons and nuclei, $m_j$ and $z_j$ are mass and charge for particle $j$, respectively, and $\hat{V}$ represents the Coulomb potential of all charged particles. The total dipole operator of the matter is $\hat{\boldsymbol \mu}=\sum_{j}z_{j}x_{j}$. In addition, $\hat{\bf x}\equiv\{\hat{\bf x}_{j}\}=\{\hat{\bf R},\hat{\bf r}\}$ with $\hat{\bf R}$ and $\hat{\bf r}$ representing the nuclear and electronic coordinates, respectively, $\hat{\bf p}\equiv\{\hat{\bf p}_{\bf R},\hat{\bf p}_{\bf r}\}\equiv\{\hat{\bf p}_{j}\}$ is the canonical momentum operator, such that $\hat{\bf p}_j=-i\hbar{\boldsymbol\nabla}_{j}$. The cavity photon field Hamiltonian under the single mode assumption is expressed as $\hat{H}_\mathrm{ph}=\hbar\omega_\mathrm{c}\big(\hat{a}^{\dagger}\hat{a}+\frac{1}{2}\big)=\frac{1}{2}\big(\hat{p}_\mathrm{c}^2+\omega_\mathrm{c}^2 \hat{q}_\mathrm{c}^2\big)$, where $\omega_\mathrm{c}$ is the frequency of the mode in the cavity,  $\hat{a}^{\dagger}$ and $\hat{a}$ are the photonic creation and annihilation operators, and  $\hat{q}_\mathrm{c} = \sqrt{\hbar/2\omega_\mathrm{c}}(\hat{a}^{\dagger} + \hat{a})$ and $\hat{p}_\mathrm{c} = i\sqrt{\hbar\omega_\mathrm{c}/2}(\hat{a}^{\dagger} - \hat{a})$ are the photonic coordinate and momentum operators, respectively. Choosing the Coulomb gauge, ${\boldsymbol \nabla} \cdot \hat{\bf A}=0$, the vector potential becomes purely transverse as $\hat{\bf A} = \hat{{\bf A}}_\perp$. Under the long-wavelength approximation, $\hat{\bf A} = {\bf A}_{0}\big(\hat{a}+\hat{a}^{\dagger}\big)={\bf A}_{0}\sqrt{{2\omega_\mathrm{c}}/{\hbar}}~\hat{q}_\mathrm{c}$ for a Fabry-P\'erot cavity, where ${\bf A}_{0}=\sqrt{\hbar/2 \omega_\mathrm{c} \varepsilon_{0}  \mathcal{V}}\cdot{\bf e}$, with {$\mathcal{V}$} as the quantization volume inside the cavity, $\varepsilon_0$ as the permittivity, and $\hat{{\bf e}}$ as the unit vector of the field polarization. 

Using the Power-Zienau-Woolley (PZW) ~\cite{PZW,Cohen-Tannoudji} gauge transformation operator $\hat{U}=\exp \big[-\frac{i}{\hbar}\hat{\boldsymbol \mu}\cdot{\bf A}_{0}\big(\hat{a}+\hat{a}^{\dagger}\big)\big]$ as well as an unitary phase transformation operator $\hat{U}_{\phi}=\exp[-i\frac{\pi}{2}\hat{a}^{\dagger}\hat{a}]$, the Pauli-Fierz (PF) non-relativistic QED Hamiltonian~\cite{Rubio2018JPB,Christian2020} $\hat{H}_\mathrm{PF}=\hat{U}_{\phi}\hat{U}\hat{H}_\mathrm{C}\hat{U}^{\dagger}\hat{U}^{\dagger}_{\phi}$ is obtained as follows
\begin{equation}\label{eqn:pf}
\hat{H}_\mathrm{PF}=\hat{H}_\mathrm{M}+\frac{1}{2}\hat{p}_\mathrm{c}^2 + \frac{1}{2}\omega_\mathrm{c}^2\big(\hat{q}_\mathrm{c} + \sqrt{\frac{2}{\hbar\omega_\mathrm{c}}}\hat{\boldsymbol\mu}\cdot{\bf A}_{0}\big)^2,
\end{equation}
where the matter Hamiltonian is $\hat{H}_\mathrm{M}=\hat{T}_{\bf R}+\hat{T}_{\bf r}+\hat{V}$, with $\hat{T}_{\bf R}$ and $\hat{T}_{\bf r}$ representing the nuclear and electronic kinetic energy, respectively. The presence of the dipole self-energy (DSE) term $\frac{\omega_\mathrm{c}}{\hbar}(\hat{\boldsymbol\mu}\cdot{\bf A}_{0})^2$ in Eq.~\ref{eqn:pf} is necessary in order to have a gauge invariant Hamiltonian~\cite{Rubio2018JPB, Christian2020} and it has shown to be crucial for an accurate description of light-matter interactions under the dipole gauge~\cite{Rubio2018JPB, Christian2020, Rabl2018PRA2}. We further assume that the dipole of the matter is oriented in the field polarization direction, such that $\hat{\boldsymbol\mu}\cdot{\bf A}_{0}=\hat{\mu}\cdot{A}_{0}$. We acknowledge that this will not be the situation of the recent VSC experiments~\cite{Ebbesen2019}, where the molecules in the solution phase should be isotropically disordered. That said, it is certainly possible to have chemical reactions in anisotropic solvents like liquid crystals~\cite{Lilichenko2003,KatoNRM2017} inside an optical cavity~\cite{KokhanchikPRB2021} More detailed discussions are provided in the Supplemental Materials.

We are interested in electronically adiabatic reactions, thus we only consider the electronic ground state of the system defined as $(\hat{H}_\mathrm{M}-\hat{T}_{\bf R})|\Psi_g\rangle=E_{g}({\bf R})|\Psi_g\rangle$, with $E_{g}({\bf R})$ as the ground adiabatic potential. Projecting $\hat{H}_\mathrm{M}$ and $\hat{\mu}$ in the ground electronic state with $\hat{\mathcal P}=|\Psi_g\rangle\langle \Psi_g|$, we obtain the following model Hamiltonian
\begin{equation}\label{eqn:model}
\hat{\mathcal H}^{g}_\mathrm{PF}=\frac{\hat{\bf P}^2}{2}+E_{g}({\bf R})+\frac{\hat{p}_\mathrm{c}^2}{2}+
\frac{1}{2}\omega_\mathrm{c}^2\Big(\hat{q}_\mathrm{c} + \sqrt{\frac{2}{\hbar\omega_\mathrm{c}}}  A_{0}\cdot {\mu}_{g}({\bf R}) \Big)^2, 
\end{equation}
where ${\mu}_{g}({\bf R})=\langle \Psi_g|\hat{\mu}|\Psi_g\rangle$. Note that projecting $\hat{\mu}$ inside the dipole self-energy term is the accurate matter state truncation scheme for the dipole-gauge Hamiltonian~\cite{Rabl2018PRA2,Taylor2020} because it ensures that all operators are properly confined in the same truncated electronic subspace $\hat{\mathcal P}=|\Psi_g\rangle\langle \Psi_g|$ in order to generate consistent results compared to the full Hamiltonian~\footnote{Indeed, if $\hat{\mathcal I}=\hat{\mathcal P}+\hat{\mathcal Q}$ represents the identity of the full electronic Hilbert space, then $\hat{\mathcal P}\hat{\mu}\hat{\mathcal P}$ is properly confined in the subspace $\hat{\mathcal P}$, whereas $\hat{\mathcal P}\hat{\mu}^2\hat{\mathcal P}=\hat{\mathcal P}\hat{\mu}(\hat{\mathcal P}+\hat{\mathcal Q})\hat{\mu}\hat{\mathcal P}$ contains the terms outside the subspace $\hat{\mathcal P}$. More numerical evidence can be found in Fig.~S2 of the Supplemental Materials in Ref.~\cite{Taylor2020}.}

In this letter, we consider a model system for $E_{g}(\bf R)$, where ${\bf R}=\{R_0, R_{1},...R_{N}\}$ include a reactive molecule with a single reaction coordinate $R_{0}$ as well as $N$ solvent DOFs $R_{i}$ (where $i\in[1,N]$) that couples to $R_{0}$. Using the typical Caldeira-Leggett~\cite{Caldeira81} system-bath Hamiltonian, $E_{g}(\bf R)$ is modeled as
\begin{equation}
E_{g}({\bf R})=U_{0}(R_{0})+\sum_{i=1}^{N}\frac{1}{2}\omega^2_{i}\big(R_{i}+\frac{c_{i}}{\omega^2_{i}}R_{0}\big)^2.
\end{equation}
The solute molecule is modeled as a double-well potential $U_{0}(R_{0}) = a\cdot R_{0}^{4} - b\cdot R_{0}^{2}$ as shown in Fig.~\ref{Model}(b) (solid black line) to represent the chemical reaction along $R_{0}$, and the details of the parameters are provided in the Supplemental Information. At the top of the barrier $R_0=R^{\ddagger}_{0}$, $U_{0}(R_{0}) \approx - \frac{1}{2} \omega_{\ddagger}^{2} (R_{0}-R^{\ddagger}_{0})^{2}$ as indicated by the red dashed line in Fig.~\ref{Model}(b), where $\omega_{\ddagger} = ~1048$ cm$^{-1}$ is the top of the barrier frequency. Further, the total dipole of the system is ${\mu}_{g}({\bf R})=\sum_{i=1}^{N}\mu_{i}(R_{i})$, and we assume that $\mu_0(R_0)=0$. Note that because the coupling strength between the cavity mode to individual molecules is very weak under the collective coupling regime; thus, the result of this letter does not change if $\mu_{0}(R_{0})\neq 0$. We further simplify our consideration of the solute-solvent coupling as $c_{i}=c_\mathrm{s}$ and $\omega_{i}=\omega_\mathrm{s} = 1400$ cm$^{-1}$. The amplitude of the solvent-friction is $\lambda=\sum_{i=1}^{N}\frac{c^2_{i}}{\omega^2_{i}}={Nc^{2}_\mathrm{s}}/{\omega^{2}_\mathrm{s}}$. In this work, we keep $\lambda$ as a constant throughout.  This means that as we increase $N$, the corresponding $c_\mathrm{s}$ will be decreased by $1/\sqrt{N}$. In particular, we use $\lambda=3.84\times10^{-4}$ a.u., and for $N=2500$, which corresponds to the solvent-solute coupling as $c_\mathrm{s}=2.5\times 10^{-6}$ a.u. Further generalization of the solvent-solute coupling with an arbitrary spectral density is possible~\cite{HughesJCP2009}, with details provided in the Supplemental Material.

Treating both ${\bf R}$ and $q_\mathrm{c}$ in Eq.~\ref{eqn:model} on an equally classical footing~\cite{Galego2019,TaoJCP2020,Jorge2020,Li2021,li2021collective}, one can express the reaction rate constant as follows~\cite{FrenkelSmit, Miller83, ChandlerBook}
\begin{equation}
  \label{rate}
k = \lim_{t\to t_\mathrm{p}}\kappa (t)\cdot k_\mathrm{TST},
\end{equation}
where $k_\mathrm{TST}$ is the Transition State Theory (TST) rate constant, $t_\mathrm{p}$ refers to the plateau time of the transmission coefficient $\kappa(t)$. The transmission coefficient captures the dynamical recrossing effects through the flux-side correlation function formalism~\cite{FrenkelSmit, Miller83, ChandlerBook}
\begin{equation}\label{kappa}
  \kappa (t) = \frac{\langle {\mathcal F}(0) \cdot
  {h}[R_{0}(t)-R_{0}^{\ddagger}]\rangle}{\langle {\mathcal F}(0)
  \cdot { h}[\dot{R}_{0}^{\ddagger}(0)]\rangle},
\end{equation}
where ${h}[R_{0}-R_{0}^{\ddagger}]$ is the Heaviside function of the reaction coordinate $R_{0}$, with the dividing surface $R_{0}^{\ddagger}$ that separates the reactant and the product regions (for the model system studied here, $R_{0}^{\ddagger}=0$), the flux function ${\mathcal F}(t)=\dot{h}(t)=\delta[R_{0}(t)-R_{0}^{\ddagger}]\cdot \dot{R}_{0}(t)$ measures the reactive flux across the dividing surface (with $\delta(R)$ as the Dirac delta function), $\dot{R}_{\ddagger}(0)$ represents the initial velocity of the nuclei on the dividing surface, and $\langle ...\rangle$ represents the canonical ensemble average with the constrain on the dividing surface enforced by $\delta[R(t)-R_{0}^{\ddagger}]$ inside ${\mathcal F}(t)$. 

It has been shown that the classical potential of mean force (free energy profile) is invariant under the change of light-matter coupling strength or photon frequency~\cite{TaoJCP2020}. Other theoretical investigations based on a simple TST analysis~\footnote{When the DSE is explicitly considered, the barrier height $E^\ddagger$ on the CBO surface remains invariant to changes of the light-matter coupling strength or the photon frequency. This is because the equilibrium position along the photonic coordinate $q_\mathrm{c}$ is $q^{0}_\mathrm{c}({\bf R})=-\sqrt{\frac{2}{\hbar\omega_\mathrm{c}}}  A_{0}\cdot {\mu}_{g}({\bf R})$ for all possible ${\bf R}$. Thus, the last term in Eq.~\ref{eqn:model} is always 0 for the reactant well or the transition state on the CBO surface. This explains why one cannot observe any effects from a simple TST analysis when treating $q_\mathrm{c}$ classically.} also suggest no significant change of the reaction rate constant~\cite{Zhdanov2020, Jorge2020}. Thus, it is reasonable to conjecture that the VSC modification of the rate constant is {\it purely dynamical}~\cite{Li2021} and completely dictated by $\kappa(t)$. Based on this, we have demonstrated the cavity frequency dependence of the VSC modification of $\kappa$ for a single molecule coupled to the cavity ~\cite{Li2021}. In this work, we consider the scenario in which such cavity modification can also be observed in the collective coupling regime. 

We numerically compute $\kappa(t)$ using the flux-side correlation function formalism in Eq.~\ref{kappa}, with the details provided in the Supplemental Materials. On the other hand, $\kappa=\lim_{t\to t_\mathrm{p}}\kappa(t)$ can also be obtained using the Grote-Hynes (GH) theory~\cite{Grote1980,Gertner1989,Peter1990,Hanggi1982,Carmeli1984,Tucker1991} through the multi-dimensional transition-state treatment~\cite{Peter1990, Eyring1935, Slater1956, Pollak1986}. The transmission coefficient using the GH theory is $\kappa_\mathrm{GH} = {{\sqrt{-(\Omega^{\ddagger}_{-})^2}}/\omega_\mathrm{\ddagger}}$,
where ${\Omega^{\ddagger}_{-}}$ is the unstable imaginary normal-mode frequency at the dividing surface $R_{0}=R^{\ddagger}_{0}$. To obtain ${\Omega^{\ddagger}_{-}}$, we further approximate the dipole of the $i_\mathrm{th}$ solvent molecule as $\mu_{i}(R_{i}) \approx \mu_{0} + \mu' R_{i}$, and define the collective bright mode of the solvent~\cite{Jorge2020} as $R_\mathrm{B} = \frac{1}{\sqrt{N}}\sum_{i=1}^{N} R_{i}$. The QED Hamiltonian ${\mathcal H}^{g}_\mathrm{PF}$ can be shown to have three coupled modes, ${\bf  x}=\{R_{0},R_\mathrm{B},q_\mathrm{c}\}$, whereas the rest of normal modes (commonly referred to as the dark modes~\cite{Jorge2020}) are completely decoupled. At the dividing surface $R_0=R_0^{\ddagger}$, the Hessian matrix in the 3-mode ${\bf x}$ subspace is
\begin{align}\label{Hessian}
\mathcal{H}_{\bf x}\equiv \frac{\partial^{2}\mathcal{H}^{g}_\mathrm{PF}}{\partial x_{i}\partial x_{j}}= \begin{bmatrix}
-\omega^2_{\ddagger} + N\frac{c_s^2}{\omega_{s}^{2}}  &  \sqrt{N} c_{s}  & 0\\
\sqrt{N} c_{s}  &  \omega^2_{s} + N\frac{\mathcal{C}^{2}}{\omega_{c}^{2}} & \sqrt{N}\mathcal{C}\\
0 & \sqrt{N}\mathcal{C}  & \omega^2_{c} \\
\end{bmatrix},
\end{align}
where $\mathcal{C} = A_{0} \mu' \sqrt{\frac{2\omega_{c}^{3}}{\hbar}}$ and $N$ is number of the solvent DOFs. A detailed derivation of Eq.~\ref{Hessian} is provided in the Supplemental Material. The imaginary frequency ${\Omega^{\ddagger}_{-}}$ can be obtained by diagonalizing $\mathcal{H}_{\bf x}$ in Eq.~\ref{Hessian}. The key to the emerging collective VSC effects is the $\sqrt{N} c_{s}$ term as well as $\sqrt{N} \mathcal{C}$ in the above Hessian matrix, which does not exist if one ignores dipole self-energies or considering $N$ solute molecules (with potential $U_{0}$) coupled to the cavity~\cite{Jorge2020}.

While ${\Omega^{\ddagger}_{-}}$ can be easily computed by numerically diagonalizing Eq.~\ref{Hessian}, a simple and concise analytical expression for ${\Omega^{\ddagger}_{-}}$ is not readily available. Nonetheless, we find (see Supplemental Material) an approximate expression of ${\Omega^{\ddagger}_{-}}$ as follows
\begin{align}\label{approximated}
\Omega^{\ddagger}_{-} \approx \Bigg[\frac{1}{ 2}(\tilde{\omega}^2_{\ddagger} -\omega^2_{c}) + \frac{1}{ 2} \sqrt{(\omega^2_{c} +\tilde{\omega}^2_{\ddagger} )^{2} + 4N\sin^{2}\Theta_{\ddagger}\mathcal{C}^{2}}\Bigg]^{1/2}
\end{align}
where $-\tilde{\omega}^2_{\ddagger} = \frac{1}{2}(-\omega^2_{\ddagger} + N\frac{c_s^2}{\omega_{s}^{2}} + \omega^2_{s} + N\frac{\mathcal{C}^{2}}{\omega_{c}^{2}}) - \frac{1}{2} \sqrt{(\omega^2_{\ddagger} - N\frac{c_s^2}{\omega_{s}^{2}} + \omega^2_{s} + N\frac{\mathcal{C}^{2}}{\omega_{c}^{2}})^{2} + 4Nc_s^2}$ and $\Theta_{\ddagger} = \frac{1}{2}\tan^{-1}[{2\sqrt{N}c_{s}/({-\omega^2_{\ddagger} + N\frac{c_s^2}{\omega_{s}^{2}} - \omega^2_{s} - N\frac{\mathcal{C}^{2}}{\omega_{c}^{2}}}})]$. It is interesting to note that Eq.~\ref{approximated} has a similar (but not identical) structure  as the case of a single molecule coupled to cavity ~\cite{Li2021}. That said, the dependence of ${\Omega^{\ddagger}_{-}}$ (or $\kappa_\mathrm{GH}$) on $\omega_{c}$ is complicated as both $\Theta_{\ddagger}$ and $\tilde{\omega}_{\ddagger}$ depends on $\omega_{c}$ in a non-trivial fashion. Nevertheless, one can clearly see from Eqs.~\ref{Hessian}-\ref{approximated} that ${\Omega^{\ddagger}_{-}}$ depends on both $\omega_\mathrm{c}$ and $N$, giving rise to the cavity frequency dependence and the collective coupling effect. Further, both ${\Omega^{\ddagger}_{-}}$ and $\kappa_\mathrm{GH}$ is also a function of $\sqrt{N} \mu'$, which is a {\it signature of the collective coupling effect}. This is similar to the collective Rabi-splitting $\hbar\Omega$ , which depends on $\sqrt{N} \mu'$ when hybridizing the solvent modes $\{R_{i}\}$ to the cavity mode $q_\mathrm{c}$ (as shown in Fig.~\ref{Model}c-d). In the Travis-Cummings model, under the resonant condition (when $\omega_{s} = \omega_{c}$), $\hbar\Omega$  is given as
\begin{align}\label{Rabi-Splitting}
\hbar \Omega = \sqrt{N}\mu' A_{0} \sqrt{2\hbar\omega_{s}} \equiv 2\eta\cdot \hbar\omega_\mathrm{c},
\end{align}
where the unitless parameter $\eta$ (defined above) characterizes the normalized light-matter coupling strength.

\begin{figure}
 \centering
  \begin{minipage}[h]{1.0\linewidth}
     \centering
     \includegraphics[width=\linewidth]{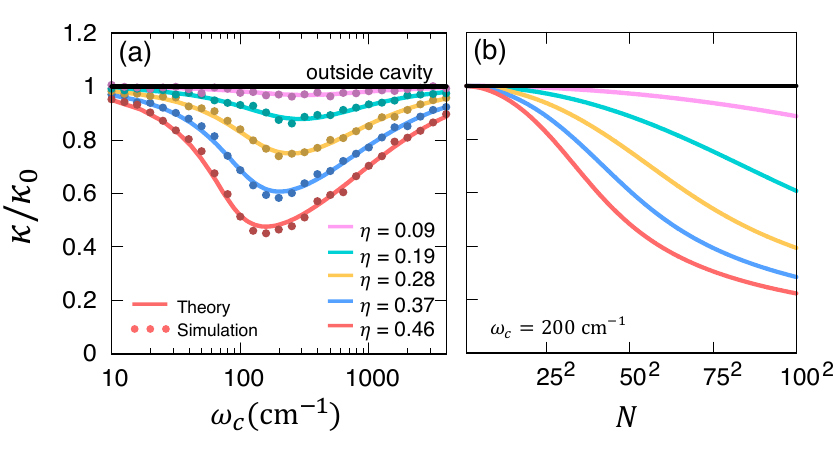}
       \end{minipage}%
   \caption{\small (a) Cavity frequency-dependent suppression of transmission coefficient $\kappa$ ($\kappa_0$ represents the outside cavity scenario) as a function of cavity photon frequency $\omega_\mathrm{c}$  with $N=2500$ solvent modes at various collective light-matter coupling strength $\eta$. (b) $\kappa$ as function of $N$ while keeping collective solvent-solute coupling $\sqrt{N}c_{s}$ the same.}
\label{noLoss}
\end{figure}
Fig.~\ref{noLoss} demonstrates both (a) the cavity frequency dependence and (b) the collective effects on modifying the reaction rate constant. Outside cavity, the transmission coefficient in the absence of light-matter interaction is  $\kappa(\eta = 0) = \kappa_{0} \approx 0.317$, which is much lower than 1 due to solvent molecules coupled to the reaction coordinate $R_{0}$ around the barrier region. 

Fig.~\ref{noLoss}a presents the results of coupling $N = 2500$ solvent DOFs to the cavity and obtaining $\kappa$ using both the GH theory (solid lines) and the direct numerical simulations (dots) using Eq.~\ref{kappa}. The results from both approaches are nearly identical. Importantly, we observe a strong dependence of $\kappa$ on the photon frequency $\hbar\omega_\mathrm{c}$, and $\kappa$ is minimized at a certain photon frequency which we refer to as $\omega_\mathrm{c}^{0}$. While we do not have a simple analytic expression of $\omega_\mathrm{c}^{0}$ (which can be in principle obtained from $\partial \kappa_\mathrm{GH} /\partial \omega_\mathrm{c}=0$), it is a function of $\omega_\mathrm{s}$, $\omega_{\ddagger}$, $c_\mathrm{s}$ and $\eta$ (see its definition in Eq.~\ref{Rabi-Splitting}). As can be seen, increasing $\eta$ results in a significant red-shift in $\omega_{c}^{0}$. Note that $\eta$ signifies collective coupling, and the individual solvent-cavity coupling is weak ($\eta/\sqrt{N} < 0.0093$). We emphasize, that the suppression of the $\kappa$ is originated from the collective dynamical caging effects, where the cavity radiation mode is effectively acting as an additional ``solvent" DOF coupled to the collective bright solvent coordinate $R_\mathrm{B}$, which in turn coupled to the reactive coordinate $R_{0}$, such that the presence of the cavity mode enhance the recrossing of the reaction coordinate and reduces the transmission coefficient.  As a result, with an increasing light-matter coupling strength, the plateau value of $\kappa(t)$ keeps decreasing, and at the same time, becoming more oscillatory (see Fig.~S3 of the Supplemental Materials). This phenomenon is well explored in the context of solvent-mediated dynamical caging effects~\cite{Baron2017, Zwan1982,Gertner1989,Peter1990,hansen2008}. More detailed discussions on the dynamical caging hypothesis of the VSC reaction can be found in Ref.~\cite{Li2021}.

Fig.~\ref{noLoss}b demonstrates the $N$-dependence of the transmission coefficient using the GH theory while keeping $\omega_\mathrm{c}$ constant. Note that to clearly identify the effect of increasing $N$ on the light-matter interactions, we have kept the solvent reorganization energy $\lambda$ a constant. We find that increasing $N$ effectively increasing the light-matter interaction strength (see Eq.~\ref{Rabi-Splitting}), leading to additional suppression of the chemical kinetics. Interestingly, unlike the $\hbar\Omega$ which linearly depends on $\sqrt{N}$ (see ~\ref{Rabi-Splitting}), $\kappa$ has a non-linear monotonic dependence of $\sqrt{N}$. This theoretical prediction  agrees with the recent VSC experiments by Ebbesen and co-workers (such as Fig.~3d in Ref.~\cite{Anoop2020}). 

Up to now, we have considered a perfect micro-cavity setup with no photon leaking. Despite the recent progress in the development of high-quality factor Fabry-P\'erot cavities, optical micro-cavities are generally leaky. The typical values of cavity losses for a Fabry-P\'erot cavity is in the range of $\Gamma_{c}=5-100$ meV~\cite{Ribeiro2018,Qiu2021,Ebbesen2019,ColesAFM2011}. In the classical Markovian limit, the cavity loss (dissipation) can be described with Langevin dynamics of the cavity radiation mode (see the Supplemental Material), with the equation of motion
\begin{align}\label{Langevin}
\ddot{q}_\mathrm{c} = - \frac{d\mathcal{H}_\mathrm{PF}^{g}}{dq_\mathrm{c}} - \Gamma_\mathrm{c} p_\mathrm{c} + {\bf F}_\mathrm{c}(t),
\end{align}
where $\Gamma_\mathrm{c}$ is cavity loss rate and ${\bf F}_\mathrm{c}(t)$ is a Gaussian random force bounded by the fluctuation-dissipation theorem through $\langle {\bf F}_\mathrm{c}(0) {\bf F}_\mathrm{c}(t)\rangle = 2\Gamma_\mathrm{c}k_\mathrm{B}T \delta(t)$. Using this approach, we can numerically compute $\kappa$ using the flux-side correlation function expression in Eq.~\ref{kappa}. Incorporating such dissipative dynamics in the GH theory makes it non-trivial to derive a simple analytic expression for $\kappa_\mathrm{GH}$ and therefore, we study the effect of the cavity loss only through direct numerical simulations.

\begin{figure}
 \centering
  \begin{minipage}[h]{1.0\linewidth}
     \centering
     \includegraphics[width=\linewidth]{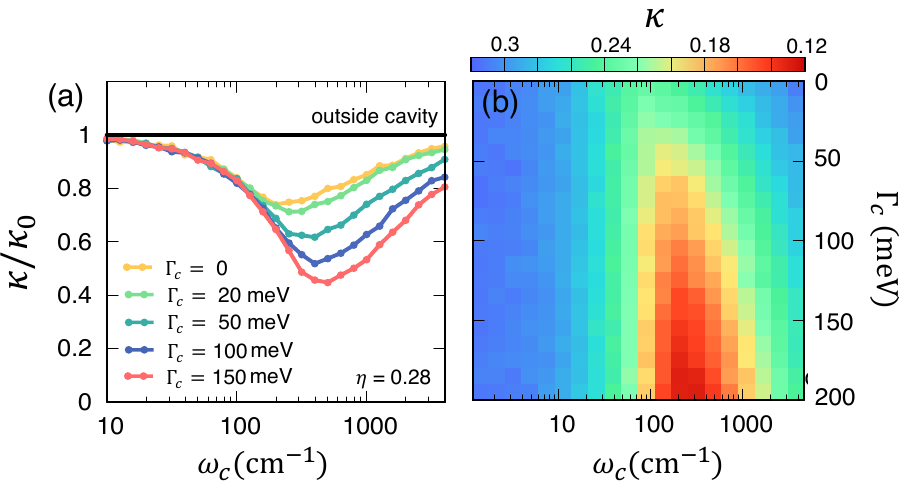}
       \end{minipage}%
   \caption{\small Transmission coefficient $\kappa$ when explicitly considering the cavity loss $\Gamma_\mathrm{c}$ at $\eta = 0.28$. (a) Cavity frequency dependence of the transmission coefficient $\kappa$ at various $\Gamma_\mathrm{c}$. (b) $\kappa$ as a function of photon frequency and cavity loss $\Gamma_\mathrm{c}$.  }
\label{Loss}
\end{figure}

Fig.~\ref{Loss} demonstrates the effect of cavity loss on $\kappa$. Here, we choose $\eta = 0.28$ and keep $N = 2500$  (with per-molecule coupling $\eta/\sqrt{N} = 0.0056$) as a constant while varying the cavity loss rate $\Gamma_\mathrm{c}$. In Fig.~\ref{Loss}a we observe that increasing $\Gamma_\mathrm{c}$ results in further suppression of the chemical rate constant while concurrently blue-shifting the maximum suppression frequency $\omega_\mathrm{c}^{0}$. It can be seen in Fig.~\ref{Loss}b that increasing $\Gamma_\mathrm{c}$ shifts $\omega_\mathrm{c}^{0}$ gradually towards to the value of $\omega_{0}$, which is the vibrational frequency of the solvent molecules. This cavity loss-assisted suppression of chemical kinetics can be attributed to the fact that the cavity loss dynamics arises when the cavity mode is coupled to other non-cavity radiation modes which act as additional dissipative baths~\cite{Feist2018PRL} (see Supplemental Material). Just like the cavity mode which acts like a dissipative bath to the collective solvent coordinate $R_\mathrm{B}$ that leads to suppression of chemical kinetics (as shown in Fig.~\ref{noLoss}), introducing an additional dissipative environment to the cavity-mode itself also leads to further suppression of the chemical rate constant. Note that when explicitly considering cavity loss dynamics, the collective coupling effects (Fig.~\ref{noLoss}) still persist, as demonstrated in the Supplementary Materials.

In conclusion, we theoretically demonstrate both the collective coupling effects and the cavity frequency-dependent modifications of the reaction rate constant in the polaritonic vibrational strong coupling regime. The model system includes the coupling between a collective solvent coordinate, the cavity radiation mode, and the reaction coordinate, giving rise to Rabi-splitting as well as cavity-modified chemical kinetics that both scales as $\sqrt{N}$, where $N$ is the number of solvent DOFs. We further demonstrate that the suppression in chemical kinetics is cavity photon-frequency dependent such that the maximal suppression occurs around a particular photon frequency $\omega^{0}_\mathrm{c}$. Further, we find a red-shift in $\omega^{0}_\mathrm{c}$  when increasing the light-matter coupling strength. In addition, we demonstrated that cavity loss significantly enhances this suppression effect and increasing cavity loss, resulting in a blue-shift of $\omega^{0}_\mathrm{c}$. Overall, when a collective solvent mode is strongly coupled to a reaction coordinate in the non-Markovian limit, the reaction coordinate becomes dynamically caged near the barrier region. We find that the cavity radiation mode as well as intrinsic cavity loss assist in dynamically caging of the reaction coordinate, leading to the suppression of the reaction rate constant if the solvent dipoles are aligned in the cavity-polarization direction. This effect operates under the collective coupling regime and survives when per-molecule (per solvent DOF) light-matter coupling is weak. We envision that the current theoretical work brings us one step closer to resolving the mysteries of VSC enabled chemistry demonstrated in recent experiments~\cite{Thomas2016, Ebbesen2019, Anoop2020, Vergauwe2019, Hirai2020-Prins, Lather2019} by demonstrating both the collective coupling effect and the cavity frequency dependent modification of the rate constant. On the other hand, we expect that the present strategy will be applicable in controlling chemical reactions in anisotropic solvents such as liquid crystals~\cite{Lilichenko2003} inside an optical cavity~\cite{KokhanchikPRB2021}.

\begin{acknowledgments}
\section{Acknowledgments}
 This work was supported by the National Science Foundation CAREER Award under Grant No. CHE-1845747, by a Cottrell Scholar award (a program by Research Corporation for Science Advancement), as well as by a University Research Award from the University of Rochester. Computing resources were provided by the Center for Integrated Research Computing (CIRC) at the University of Rochester. The authors appreciate valuable comments to the manuscript from Braden M.  Weight.
\end{acknowledgments}

%

\end{document}